# Free electron lifetime achievements in Liquid Argon Imaging TPC

**B. Baibussinov[a], M. Baldo Ceolin[b,a], E. Calligarich[c], S. Centro[b,a], K. Cieslik[a,d], C. Farnese[b,a], A. Fava[b,a], D. Gibin[b,a], A. Guglielmi[a], G. Meng[a], F. Pietropaolo[a], C. Rubbia[e,f,*], F. Varanini[b,a] and S. Ventura[a]**

[a] *INFN, Sezione di Padova*
*via Marzolo 8, I-35131 Padova, Italy*
[b] *Dipartimento di Fisica, Università di Padova*
*via Marzolo 8, I-35131 Padova, Italy*
[c] *INFN, Sezione di Pavia*
*via Bassi 6, I-27100 Pavia, Italy*
[d] *Instytut Fizyki Jadrowej PAN*
*Krakov, Poland*
[e] *Laboratori Nazionali del Gran Sasso dell'INFN*
*I-67010 Assergi (AQ), Italy*
[f] *CERN, European Laboratory for Particle Physics*
*CH-1211 Geneve 23, Switzerland*

*E-mail*: Carlo.Rubbia@cern.ch

ABSTRACT: A key feature for the success of the liquid Argon imaging TPC (LAr-TPC) technology is the industrial purification against electro-negative impurities, especially Oxygen and Nitrogen remnants, which have to be continuously kept at an exceptionally low level by filtering and recirculating liquid Argon. Improved purification techniques have been applied to a 120 liters LAr-TPC test facility in the INFN-LNL laboratory. Through-going muon tracks have been used to determine the free electron lifetime in liquid Argon against electro-negative impurities. The short path length here observed (30 cm) is compensated by the high accuracy in the observation of the specific ionization of cosmic ray muons at sea level as a function of the drift distance. A free electron lifetime of $\tau \sim (21.4^{+7.3}_{-4.3})$ ms, namely > 15.8 ms at 90 % C.L. has been observed over several weeks under stable conditions, corresponding to a residual Oxygen equivalent of ≈ 15 ppt (part per trillion). At 500 V/cm, the free electron speed is 1.5 mm/μs. In a LAr-TPC a free electron lifetime in excess of 15 ms corresponds for instance to an attenuation of less than 15 % after a drift path of 5 m, opening the way to the operation of the LAr-TPC with exceptionally long drift distances.

* Corresponding author.

## Contents

## 1. Introduction

Bubble chambers have had a major role in neutrino physics: for instance the well known Gargamelle bubble chamber, in spite of its relative small sensitive mass compared to other electronic calorimetric detectors has contributed in an essential way to the discovery of Neutral Currents [1] and of the Electro Weak components of the Standard Model. In the future the use of a totally sensitive imaging detector may have a unique role for instance in neutrino physics, proton decay and the search for WIMP like events coming from dark matter. These future detectors however do require additional features: they must be continuously sensitive rather than pulsed, they may obey to the strict safety requirements in underground laboratories and they must be capable of reaching a sensitive mass of many tens of thousands of tons (as a comparison, the Gargamelle chamber had a sensitive mass of 3.14 tons of Freon).

Argon, with its 0.9 % content in air, is the only noble gas, which has today a vast industrial utilization, since it can be easily produced in very large quantities and with a remarkably low cost. Its main production is part of the liquefaction of dry air and it can be easily separated by fractional distillation for instance from Nitrogen and Oxygen.

As early as in 1977 a new method of imaging in a noble liquid was proposed [2], based on a truly novel concept of free electron drift as a function of the time, to observe the true "image" of the track with a resolution of the order of 1 $mm^3$, thus extending to a liquid the idea of the TPC already described for a gas, originally proposed by G. Charpak et al. [3]. The method has been extensively studied by ICARUS over two decades, firstly with small laboratory-like detectors and later with much larger detectors developed with the help of industry. At present ICARUS T600 [4], the largest operated liquid Argon (LAr) detector, a large "swimming pool" arrangement with a volume in excess of 600 tons and 1.5 m long electron drift path is now finally installed deep underground in Hall B of the LNGS laboratory (Assergi, Italy), to be exposed to cosmic neutrinos and to the neutrino beam from CERN (CNGS2).

## 2. Key features of LAr imaging

Some of the main features of the LAr-TPC are hereby briefly summarized. The minimum ionization particle yield is of about 8400 electrons/mm, reduced to about 4600 e/mm at 500 V/cm because of columnar recombination. With modern front-end j-FET amplifier, a few mm long minimum ionization particle track is well above noise background. At 500 V/cm, the drift speed is 1.5 m/ms: a 5 m drift length corresponds to a drift time of 3.3 ms. The intrinsic bubble size (r.m.s diffusion) is given by $\sigma_D$ [*mm*] $\approx 0.9 \sqrt{(T_D [ms])}$. The values for 5 m drift are $\langle\sigma_D\rangle \approx 1.1$ mm and $\sigma_{max} \approx 1.3$ mm, tiny with respect to the envisaged wire pitch ($\geq$ 2mm). Therefore even at long drift distances, the small size of the "bubble" is preserved.

According to the kinetic theory, the densities (particle/cm$^3$) of the (gaseous) contaminating impurities are the same in the liquid and in the gas regions. However, since the density of the (cold) Argon gas (GAr) is some 200 times smaller than the one of the liquid, the drift speed of diffusion of impurities is very slow, << 1 m/hour and the establishing of the equilibrium relays mostly on convective motions rather than on diffusion. Moreover, most impurity sources are in gas phase (material degassing, possible leaks, …), hence the fractional density of impurities is much larger in gas than in liquid. Therefore the evacuation of the polluted GAr evaporating from the TPC soon after the LAr filling could efficiently improve the LAr purification process.

The present paper reports new results on the attainable drift length in a practical detector. The longest drift distance at which one can operate with free electrons is inversely proportional to the residual contamination due to several electro-negative gases, far more demanding since the free electron path in a liquid is $\approx$ 600 times shorter than in a gas. Although several polar chemicals may be present in the volume, we have used the concept of contamination Oxygen equivalent. For instance a 10 ms lifetime corresponds to 30 ppt (t = trillion!) of $O_2$ equivalent.

During 2001 a first real size operation of one half of the ICARUS T600 detector was performed on surface in Pavia. The drift length of the detector was set to 1.5 m, corresponding to a drift time of about 1 ms at the nominal field of 500 V/cm. The purity performance was of the order of 2 ms, measured both with purity monitors and muon tracks, uniform over the whole detector. Considerable progress over the last few years has permitted to reach industrial purification techniques that are capable of a much better performance reported in the present paper.

## 3. The experimental setup

The LAr-TPC detector has been realized in the framework of the ICARUS R&D activity at the INFN-LNL laboratories in Legnaro (Italy). The chamber is an improved version of the LAr-TPC used at CERN on the WANF neutrino beam in 1997-98 [5]. It is contained in a stainless steel cylindrical vessel, 60 cm diameter and 100 cm height, whose upper face is an ultra high vacuum (UHV) flange housing the feed-through for vacuum, the LAr filling and re-circulation, the high voltages and the read-out electronics. The total volume of the internal dewar is 283 liters, 120 of which are filled with LAr. The TPC consists of two vertical electrode planes, 32.6 x 32.6 cm$^2$ acting as anode and cathode, laterally delimited by 4 vetronite 29.4 x 29.4 cm$^2$ boards supporting the field-shaping electrodes). The resulting active TPC volume is 29.4 x 29.4 x 31.8 cm$^3$ corresponding to 38 kg LAr mass (Figure 1).

The read-out electrodes, forming the anode, are two parallel wire planes spaced by 3 mm: the first one (facing the drift volume) works in induction mode and is made of vertical wires while the second one collects the drifting electrons and is made of horizontal wires. Each plane is made of 96 stainless steel wires, 100 μm diameter and 3 mm pitch. The wires are soldered on a vetronite frame, which supports also the high voltage distribution and the de-coupling capacitors.

A third wire plane, commonly called grid, is inserted 3.5 mm in front of the Induction wires. It acts as an electromagnetic screen against the noise due to the HV biasing, improving the induction signal shape. The grid is electrically biased but no signal is recorded from this plane. Finally, another thin stainless steel grid has been inserted behind the Collection wire plane and put to ground.

In order to facilitate the LAr re-circulation in the TPC active volume, the cathode at the front of the chamber has been realized with a thin etched stainless steel grid (Figure 1-right).

The field-shaping electrodes consist of 30 stripes of golden copper on G10 vetronite boards: the first and the last ones are 5 mm wide, while the other 28's are 8 mm wide, all spaced by 2 mm. A high-voltage divider, made of a series of resistors (20 MOhm, 8.17 MOhm, 27 x 10 MOhm, 8.58 MOhm and 11.76 MOhm from anode to cathode), interconnects and supplies the correct voltage to each field-shaping ring in order to obtain an uniform electric field. Each resistor value has been measured at room temperature with a precision better than 1 %.

The high voltage for the drift field is brought to the cathode through a custom made cryogenic ultra high vacuum feed-through able to stand up to 30 kV. An electric field of 474 V/cm is applied in the drift volume, corresponding to - 14.8 kV on the cathode. In order to ensure a full transparency of the Grid and of the Induction plane to drifting electrons, the potential is fixed at 350 V for the Collection wires, - 100 V for the Induction wires and – 350 V for the Grid. This set of values was chosen after a careful study on the reconstructed muon tracks collected in several preliminary test runs, optimizing the electric field uniformity in the drift region between the anode and the beginning of the race-track.

The whole detector vessel is contained in an open-air stainless steel dewar, which is initially filled with commercial LAr acting as cryogenic bath for the ultrapure LAr injected in the detector vessel.

A standard sub-set of the ICARUS DAQ [4] system is used to read-out the 192 wires of the TPC. The Induction and the Collection planes are read with 3 charge-like analogue boards (CAEN-V791Q) and 3 current-like analogue boards (V791C) respectively. Hence, the charge generated by ionizing particles is measured on the collection plane by means of the signal pulse area.

The analogue signals from the preamplifiers are first multiplexed (16:1) and then digitized (with a sampling time of 400 ns) through a 10-bit Flash-type Analogue to Digital converters. Data are continuously recorded in circular memory buffers on the digital board CAEN-V789, and transferred on mass storage devices for successive off-line analysis only when a trigger occurs.

For the test run described here, the DAQ was run with a global stop signal provided by the external scintillation trigger system; the buffer size was chosen to be 1024 t-samples (one t-sample: 400 ns), sufficiently wide to record a full drift time (about 500 t-samples). An event display and a reconstruction code à la ICARUS are used to extract hit signals, filtering them from the background noise for both the Collection and the Induction views.

## 4. Purification of the LAr

The TPC vessel is filled with ultra pure LAr, following the standard ICARUS procedure [4]. First the chamber is evacuated to at least $10^{-3}$ Pa with turbo-molecular pumping system for several days, to maximize degassing of the detector materials. Then the chamber is cooled down in the LAr bath as fast as possible and is filled with commercial LAr purified by the Oxysorb/Hydrosorb™ filter.

An additional tank (30 cm diameter, 50 cm height) is connected to the TPC vessel with the aim of removing the gaseous Argon, produced by the evaporation on the warm inner detector surfaces during the first phase of the filling and possibly contaminated by degassing. The additional vessel is also evacuated to $10^{-4}$ Pa and kept immersed into a LAr bath inside an independent dewar. At the beginning of the filling procedure the detector material is warmer than LAr temperature, thus inducing a twofold effect:

1) degassing of impurities is still effective, possibly polluting the injected ultra-pure Argon;
2) the LAr injected into the cryostat is partially evaporated to cool the detector components, slowing down the filling procedure. This contributes to reduce the initial purity because more time is available for degassing at temperature higher than that of LAr.

The use of an additional vessel to extract and liquefy the possibly polluted GAr evaporating from the TPC has a double effect of trapping the impurities eventually present in the gas and speeding up the filling procedure. As a result, almost pure LAr could be obtained right after the filling. The additional vessel is disconnected soon after the completion of the filling procedure, and the ordinary recirculation initiated. Moreover, the LAr cryogenic bath, in which the detector vessel is immersed, is evacuated from the open-air dewar and the recirculation system is used to compensate for the detector heat losses. The entire LAr test facility is sketched in Figure 2.

The gaseous Argon, evaporating from the liquid in the detector vessel, is purified by the Oxysorb/Hydrosorb filter and is then condensed in a passive heat exchanger (LAr bath) and finally is re-injected at the bottom of the detector. During this recirculation process, the GAr flowing through the purifier is warmed up to room temperature, being the transfer lines not thermally insulated. Thus the heat exchanger must both cool the GAr down to the boiling point (87 K) and liquefy it; given the values of the evaporation latent heat (38.4 cal/g) and of the thermal capacitance (0.267 cal/g K), the liquefaction efficiency of the system results to be 42 % (i.e. 2.4 liters of LAr in the thermal bath are evaporated for every liter of purified and re-condensed LAr in the dewar). Since the total LAr consumption in the heat exchanger is 4.5 liters/hour, the recirculation flow results to be about 1.9 l/h. In these conditions the total amount of LAr is completely recirculated every 2.5 days.

This set-up reproduces in smaller scale the gas recirculation circuit installed on ICARUS-T600, which is in fact dimensioned to function with gas purification only. The use of the additional cryogenic container in the LNL laboratory set-up to get a very high initial purification level, has been included to replace the liquid recirculation system present in the T600 layout, which is meant to speed up the purification immediately after the LAr filling and, hopefully, for a limited time duration.

## 5. Experimental results

A precise calibration of the response of each electronic channel was performed with a dedicated test pulse circuit, in order to accurately measure the charge attenuation along tracks. For this purpose, a special board has been used, with 32 calibrated capacitors of value 3.22 pF with a tolerance of 0.05 pF and test pulse from stable generator. This special calibration board replaces the wire cables, thus allowing precise calibration of the full front-end electronic chain including the decoupling capacitor boards (Figure 3); in fact the detector input capacitance is already precisely knows due to the equal length of all wires (~30 cm) and all cables (3.5 m). Changing pulse height and rise time of the voltage step, a charge pulse in 2.5-75 fC range could be injected into each electronic channel with durations spanning a 0-4 μs time interval. Pulses in the calibration changed from ~ 10 ADC counts (slightly lower than the m.i.p. signals recorded from cosmic ray muons) to ~ 400 ADC counts. As a comparison, a typical m.i.p. signal detected by the ICARUS electronics has a duration of ~ 3 μs.

The distribution of the signal integral ("area" of the hit) of ~ 500 recorded events for each wire was fitted with a Gaussian function for each selected pulse-height and rise time values. The obtained values provide the relative charge calibration constant for each wire, which differ of about 1% (Figure 4). At the largest pulse height the signal area of each channel resulted stable better than 0.5 % level with respect to the signal duration.

The gain of each wire is affected by a 0.8 % systematic error due to the capacitances used in the calibration procedure. A 0.3 % term is added, depending on the pulse duration.

Through-going muons were triggered by external scintillation counters (~40 $cm^2$ active area) crossing most of the drift length at about 45° with respect to the vertical axis (Figure 5). Data taking for LAr purity measurement with cosmic rays started few days after the LAr filling completion, when stable cryogenic conditions, i.e. LAr uniform temperature inside the chamber volume, were reached. The run lasted about three weeks; with a trigger rate of 1/250 $s^{-1}$, a total of about 6300 events were collected. Temperature, pressure and level of LAr in the chamber were continuously monitored during data taking. Single cosmic muon track events are selected off-line in few successive steps analyzing the Collection view signals. Few simple events selection criteria have been applied to extract an isolated track sample, without large δ-rays and accompanying spurious hits, and spanning the full 27-187 μs fiducial drift region (see Table 1). As a conclusion, 1172 clean muon tracks, crossing the detector at 45°±1.6°, were selected out of the initial sample.

Table 2. Effects of the various event selections criteria on the initial sample of 6320 events.

| Selection | Rejected events | Surviving events |
|---|---|---|
| Empty events and e.m. showers | 1184 | 5136 (81.3%) |
| Noisy events | 62 | 5074 (80.3%) |
| Clusterization | 649 | 4425 (70.0%) |
| δ-rays | 2723 | 1702 (26.9%) |
| Short tracks | 530 | 1172 (18.5%) |

In the reconstruction software the track signals are recognized imposing a threshold of ~50% (5 ADC counts) of the average pulse height in the rising and falling edges, and a threshold in the hit width at ~50% (6 t-samples) of the typical FWHM of mip signals. Then a fit of the signal is performed and for each hit the wire number, the arrival time, and the pulse area over the baseline are recorded. The latter is proportional to the total charge (free electrons) induced on each collection wire.

Systematic errors associated to measurement of the signal of the single hits have different origins. A first 0.1 % contribution comes from the maximal variation of measured baseline and in particular of its possible dependence from the drift time. An absolute uncertainty on the baseline determination was estimated by comparing different extraction algorithms. Its contribution to the relative charge determination resulted to be at the 0.1 % level.

Residual non-uniformities of the electric field in the fiducial volume could generate track distortions accompanied by focusing / defocusing of the charge collected by individual wires. A 0.5 mm maximal deviation from the linear fit of the track was measured in the selected 27 to 187 μs drift time interval (Figure 6), exhibiting a residual parabolic shape, which has the effect of slightly reducing the collected charge with increasing drift time. Hence this small electric field related systematic error cannot bias the measured electron lifetime towards higher values.

Moreover, given the 7 GeV average energy of the selected 45° cosmic muons, dE/dx variation along the 30 cm track is negligible.

A first evaluation of the mean electron lifetime $\tau$ has been performed on an event-by-event basis by means of a linear fit of the logarithm of the recorded muon hit charge versus drift time. In order to reject hot spots (hits with very high energy deposition which populate the Landau tail) the method of the logarithmic truncated mean has been applied, excluding the upper 10 % of the hits (Figure 7). The resulting distribution for the full sample as a function of the logarithm of the pulse height area is shown in Figure 8, with and without the method of the logarithmically truncated mean, without the largest 10 % points showing that this choice is adequate to make the distribution gaussian. A 14 % bin-to-bin uncertainty in the charge measurement was estimated in the single event fit. The results give an estimate of the mean free electron attenuation $<1/\tau> = (0.051 \pm 0.015)$ ms$^{-1}$ ($\chi^2$ / d.o.f = 59/58). A 0.011 ms$^{-1}$ overall systematic term, due mainly to the calibration procedure previously discussed, has to be added, providing an electron lifetime estimation $\tau \sim (20.^{+12.}_{-5.})$ ms.

In a almost independent method, the scatter plot of the hit area (proportional to the charge) as a function of the drift time for the overall selected muon sample (see Table 1) has been sliced into 16 time-bins, and for each of them a representative value of the hit charge has been extracted from a Landau convoluted with a Gaussian fit of the distribution of the hit areas. The latter is included to take into account the smearing due to electronic noise, muon track length variations due to trigger angular acceptance and track-to-track dE/dx variations due to the cosmic ray energy spectrum. The statistical error on the peak determination of the Landau-fit amounts to about 0.3 %. A 0.3 % systematic uncertainty comes from the pulse calibration, since about 15 wires (each one affected by 0.8 % systematic error) contribute to the average charge in each time drift bin. As a result, the measured average charge versus the drift time shows a residual slope $1/\tau = (0.047 \pm 0.012)$ ms$^{-1}$ ($\chi^2$ / d.o.f = 11/14), corresponding to an effective electron lifetime $\tau = (21.4^{+7.3}_{-4.3})$ ms (Figure 9).

## 6. Conclusions

A 120 l LAr test facility (38 kg active mass) has been presented. It is equipped with the standard ICARUS cryogenic recirculation-purification system and with a subset of the standard ICARUS DAQ system. The apparatus has been used to perfect the purity measurement with through-going cosmic muons. The short path length used (30 cm) is compensated by the high accuracy in the observation of the specific ionization.

In the study presented the detector has been operated for about one month, collecting about three weeks of useful data, after seven days from the initial filling, necessary to optimize the system and to reach a stable cryogenic equilibrium.

Methods have been devised to accurately measure the deposited charge by through-going cosmic muons and obtain an estimate of the electron lifetime. No indication of variation of purity during the data taking has been observed, so the whole data sample of 1172 tracks has been used in the measurement. No variations of the electron lifetime during this period have been observed within the collected statistics (Figure 10).

With the adopted filling and purification procedure, the free electron lifetime has been found to be $\tau = (21.4^{+7.3}_{-4.3})$ ms as measured with the global fit method, and confirmed by the event-by-event fit evaluation $\tau \sim (20.^{+12.}_{-5.})$ ms. The result corresponds to $\approx$ 15 ppt (parts per trillion) of Oxygen equivalent, or $\approx 10^{-11}$ molecular $O_2$ impurities in Ar. The excellent purity level obtained would correspond to less than 15 % electron signal attenuation for a 5 m drift at 500 V/cm electric field, opening the way to LAr-TPC's with exceptionally long drift distances [6].

## 7. Acknowledgements

We wish to acknowledge the work of the many people of the ICARUS Collaboration, in the framework of which this work was made possible. We would like to express our warm thanks to the technical collaborators, who contributed to the construction of the detector and to its operation; in particular D. Filippi, M. Nicoletto as well as many members of the "Servizio di Meccanica" and "Servizio di Elettronica" of the Istituto Nazionale di Fisica Nucleare (INFN) "Sezione di Padova". We are indebted to C. Bellunato, D. Dequal, E. Scantamburlo and R. Zonta for their contribution in the data taking and analysis, during their thesis preparation. Special thanks go to the CRIOTEC IMPIANTI for the skill and the collaborative effort shown in the design and construction of the cryogenic facility. Finally, this work would not have been possible without the financial and technical support of our funding institutions, INFN and MIUR, Italy.

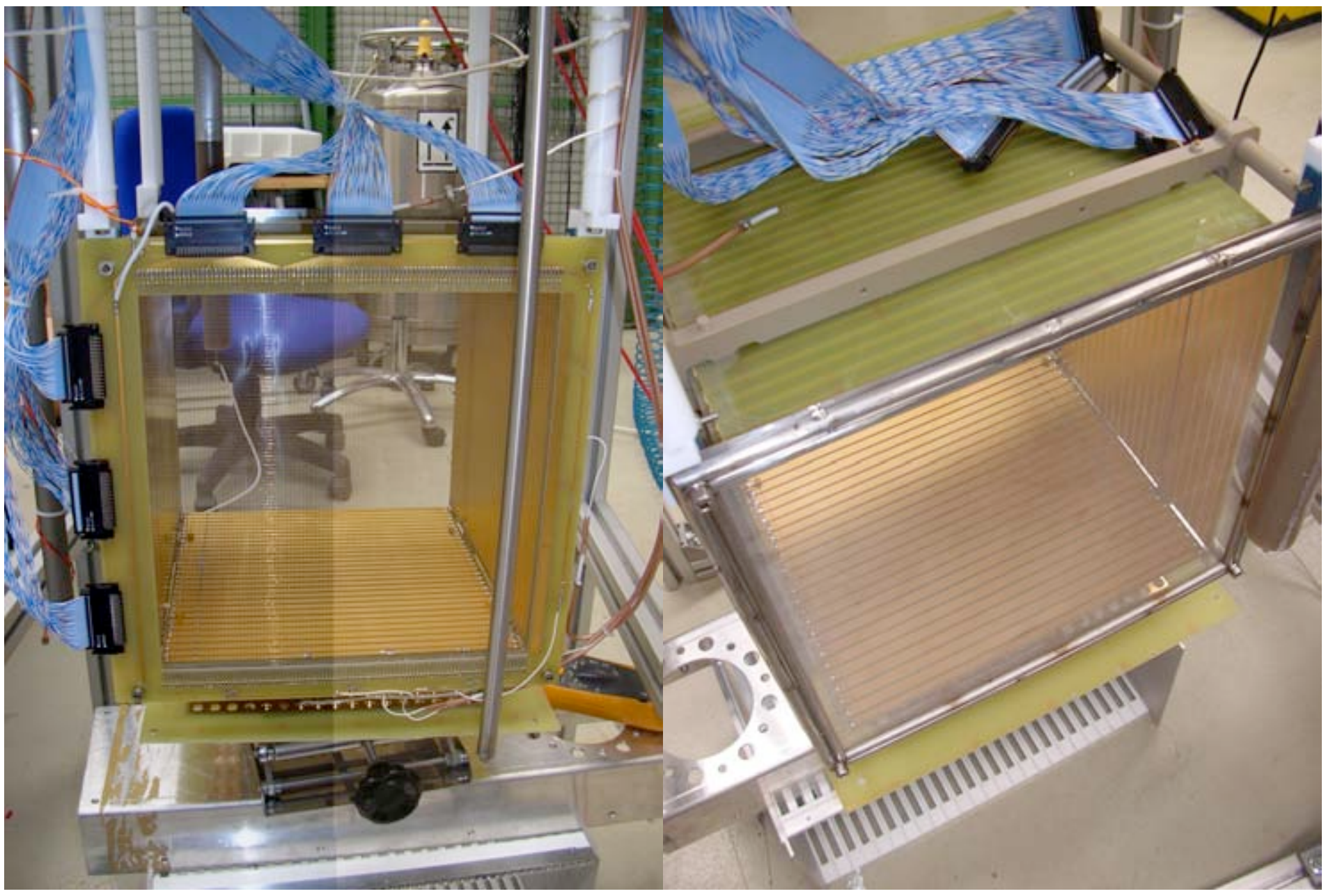

**Figure 1**. Photos of the LAr-TPC at LNL showing the anode structure (left) and the cathode (right). The field shaping electrodes are also visible.

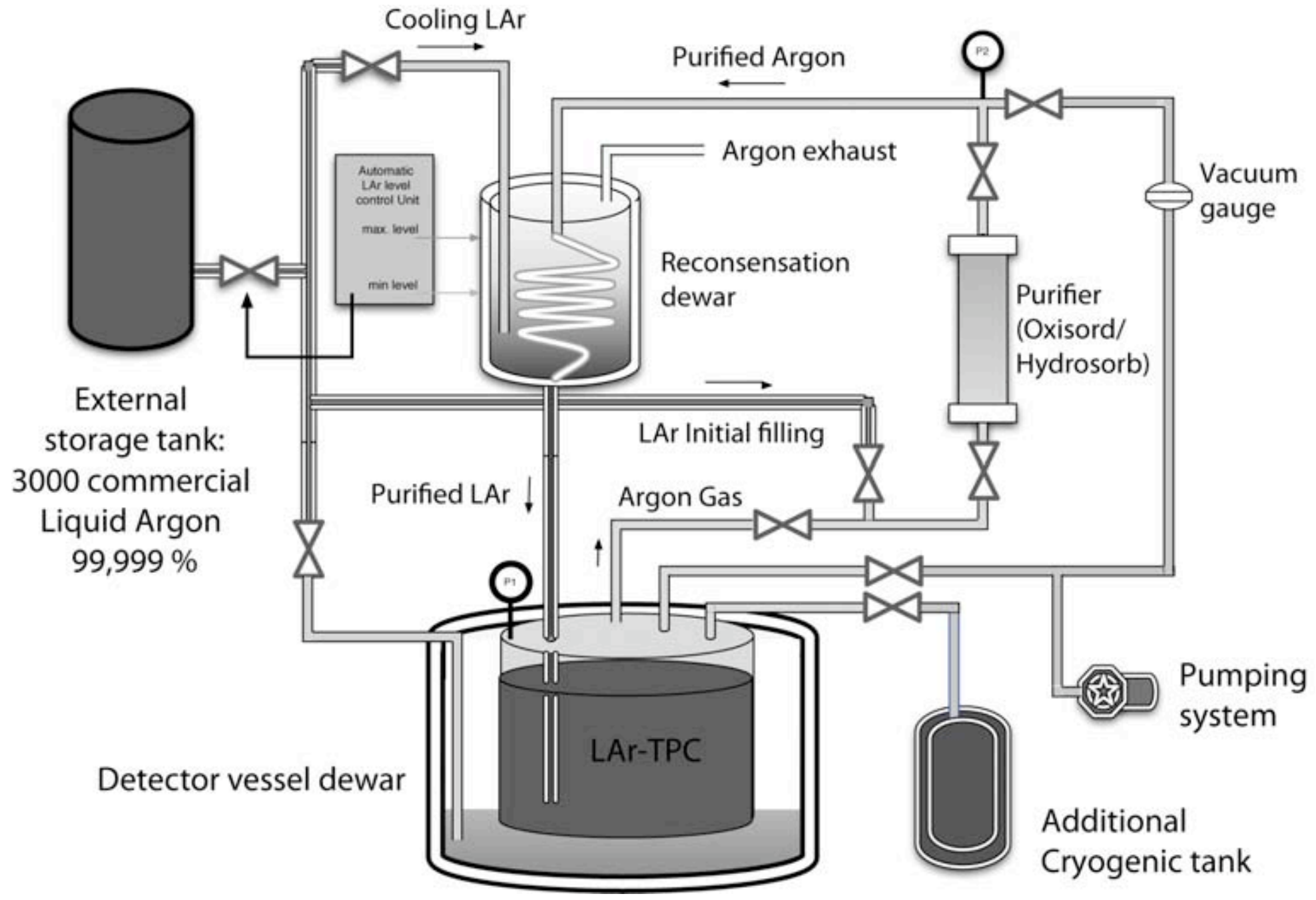

**Figure 2.** Scheme of the cryogenic circuits of the LAr-TPC facility at LNL.

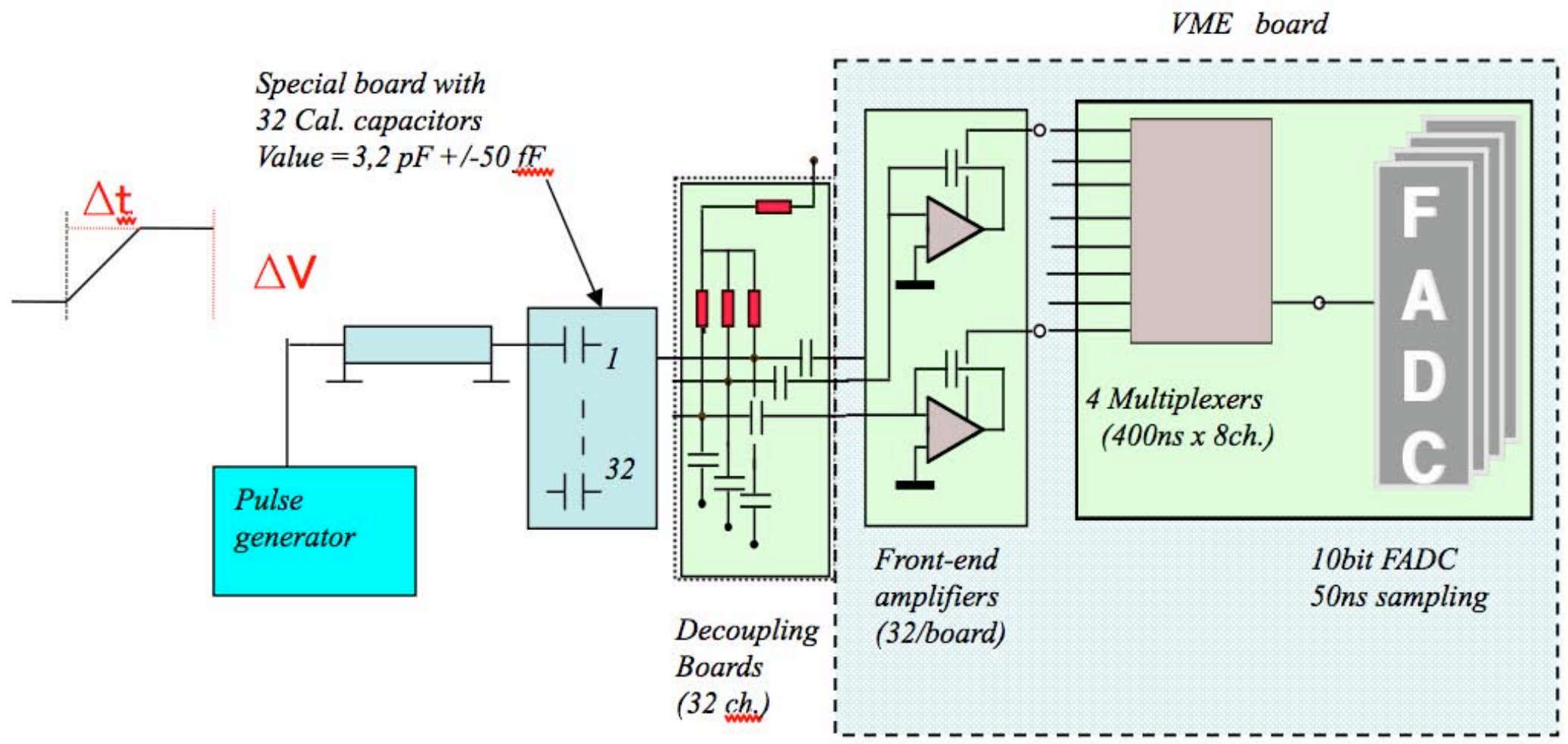

**Figure 3**. Scheme of the electronic set-up used for the signal calibration with test-pulses.

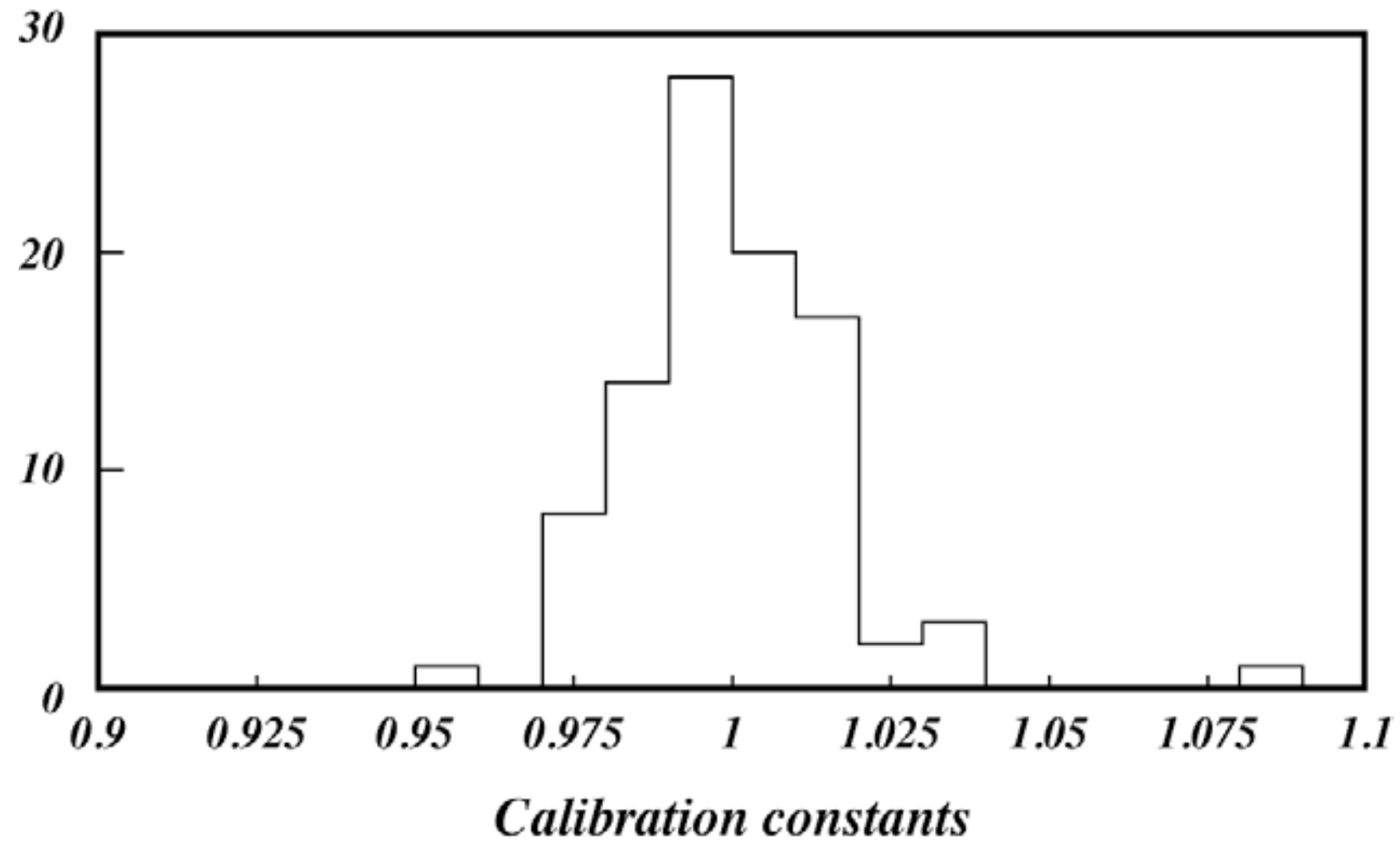

**Figure 4**. Distribution of the relative wire signal calibration constants (RMS = $1.7\ 10^{-2}$) referred to the average value.

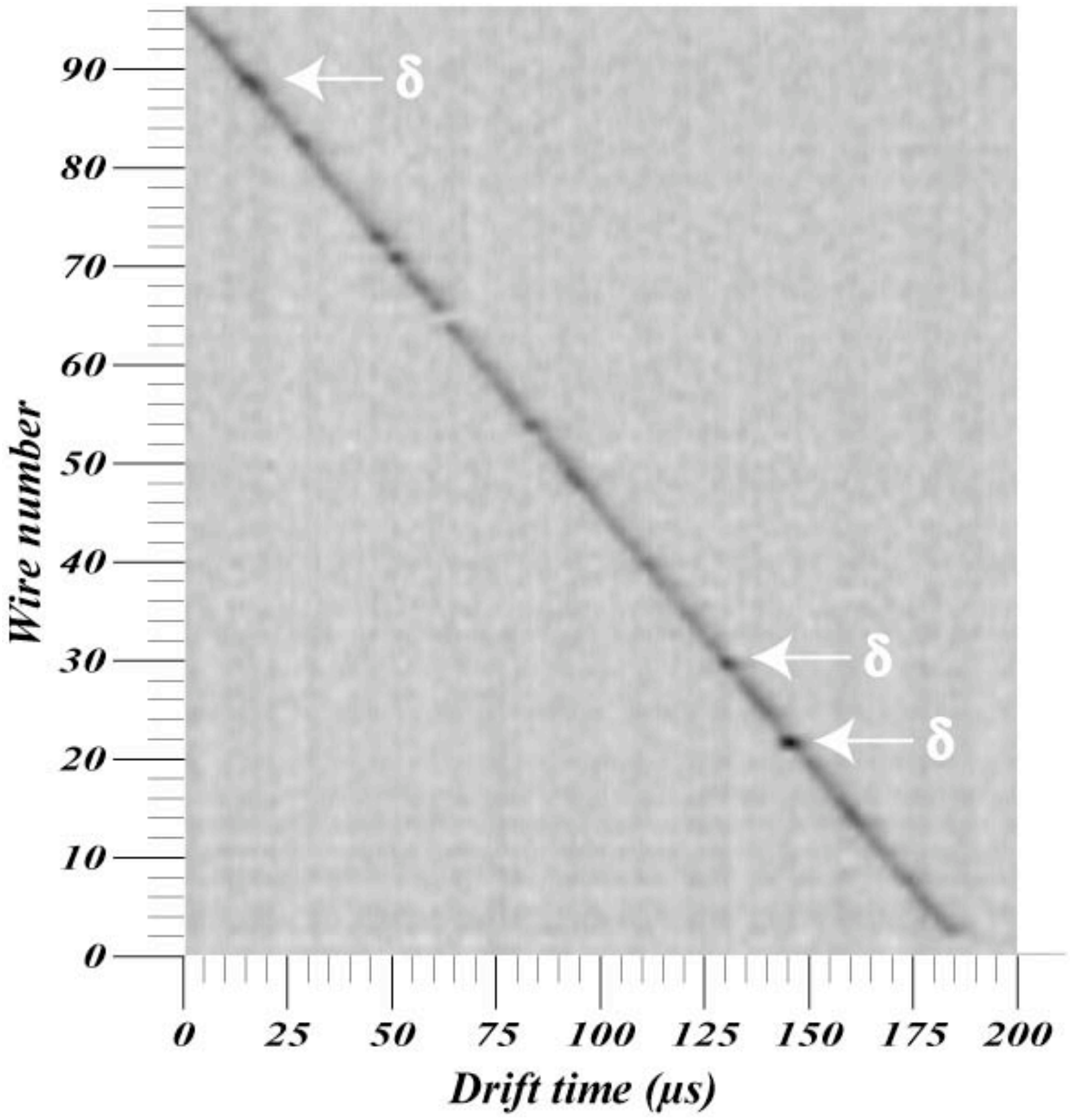

**Figure 5.** Typical muon event. The grey level is proportional to the energy deposition. Note the presence of additional energy depositions due to δ-rays.

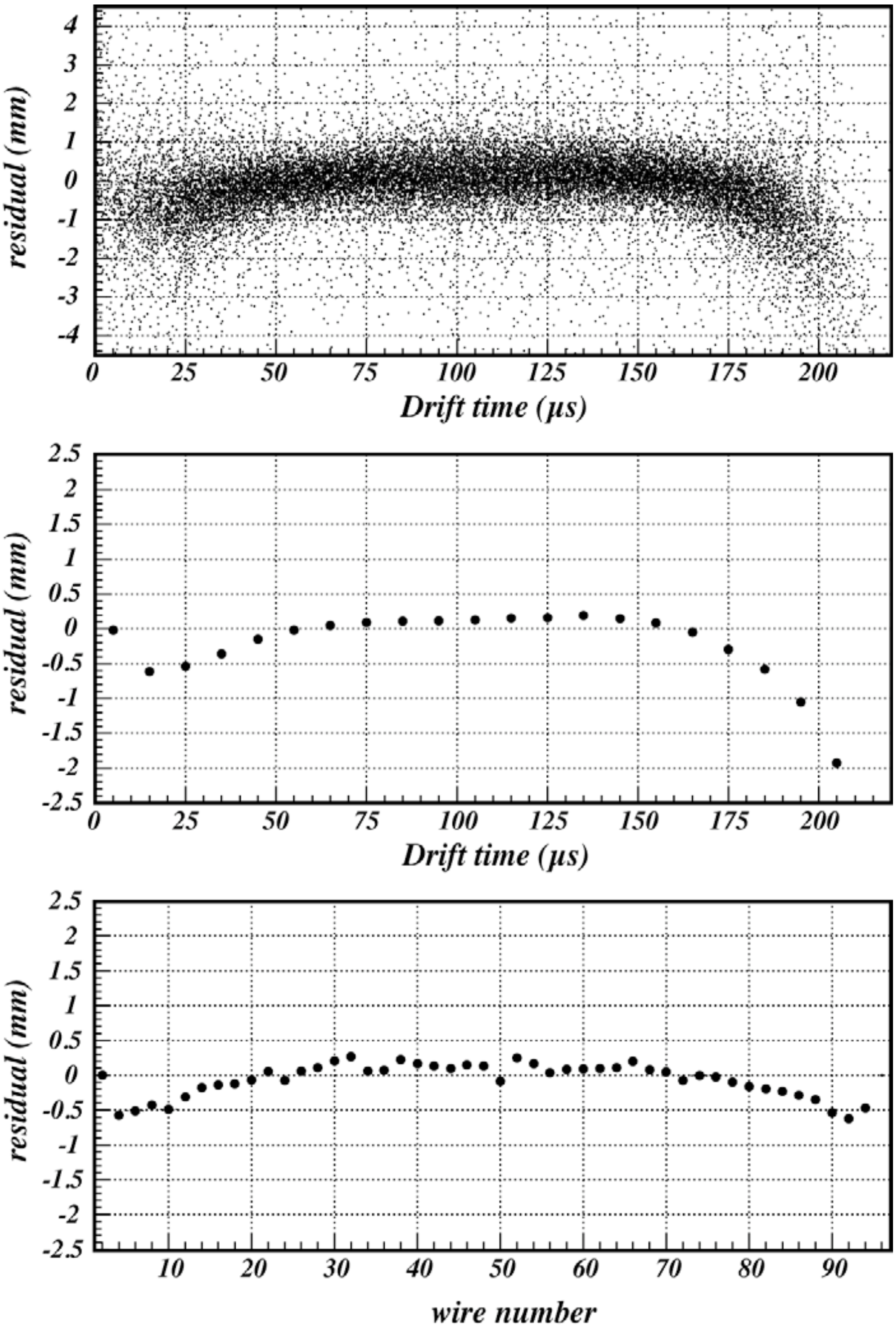

**Figure 6.** Residual distribution of the hit distance from a straight line fit of the track as a function of the drift time in the Collection view: event by event (top) and average values (center). The corresponding average values as a function of the hit wire are also shown (bottom); in this last plot the selection of the hits between 187 and 580 t-samples (namely between 27 and 187 μs) has already been performed.

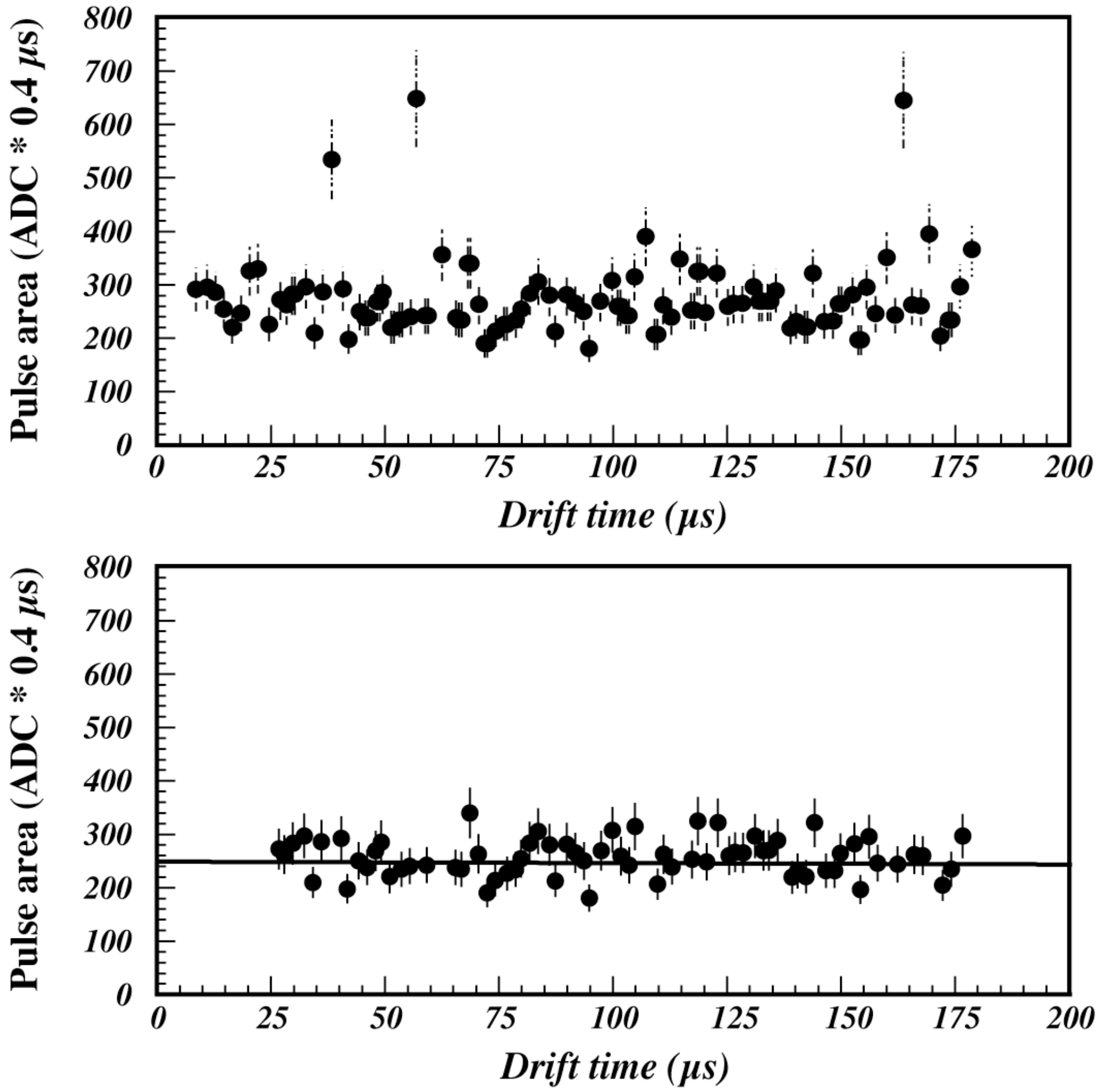

**Figure 7.** Pulse height area as a function of the drift time for a single track. *Upper plot*: raw points with the energy depositions due to δ-rays. *Lower plot*: the same track using the method of the logarithmically truncated mean, with the largest 10 % points rejected. The fitted line specifically for this event is $1/\tau = (0.11 \pm 0.39)\ \mathrm{ms}^{-1}$ ($\chi^2$ / d.o.f = 61 / 65).

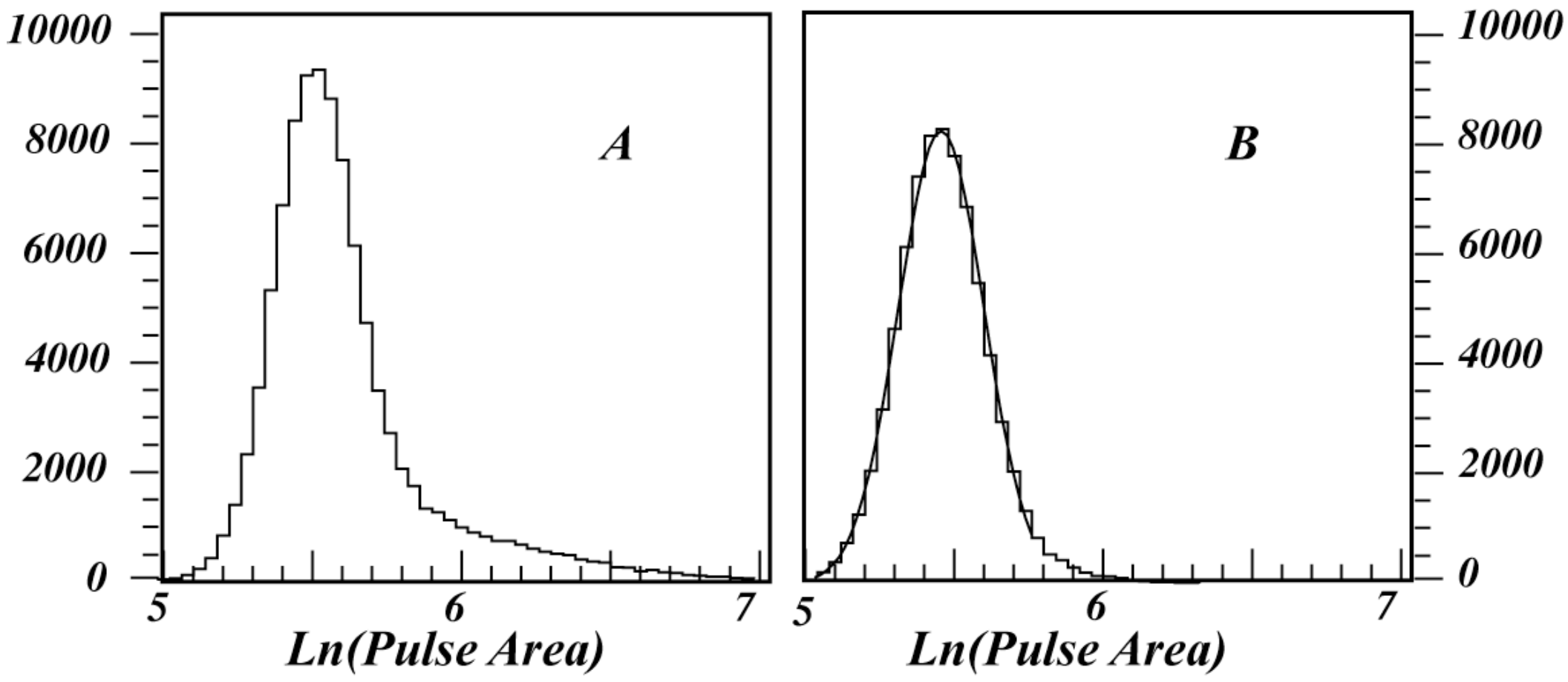

**Figure 8**. (A) Number of valid events as a function of the logarithm of the pulse height area. (B) The same as previous, but with the method of the logarithmically truncated mean, with the largest 10 % points rejected.

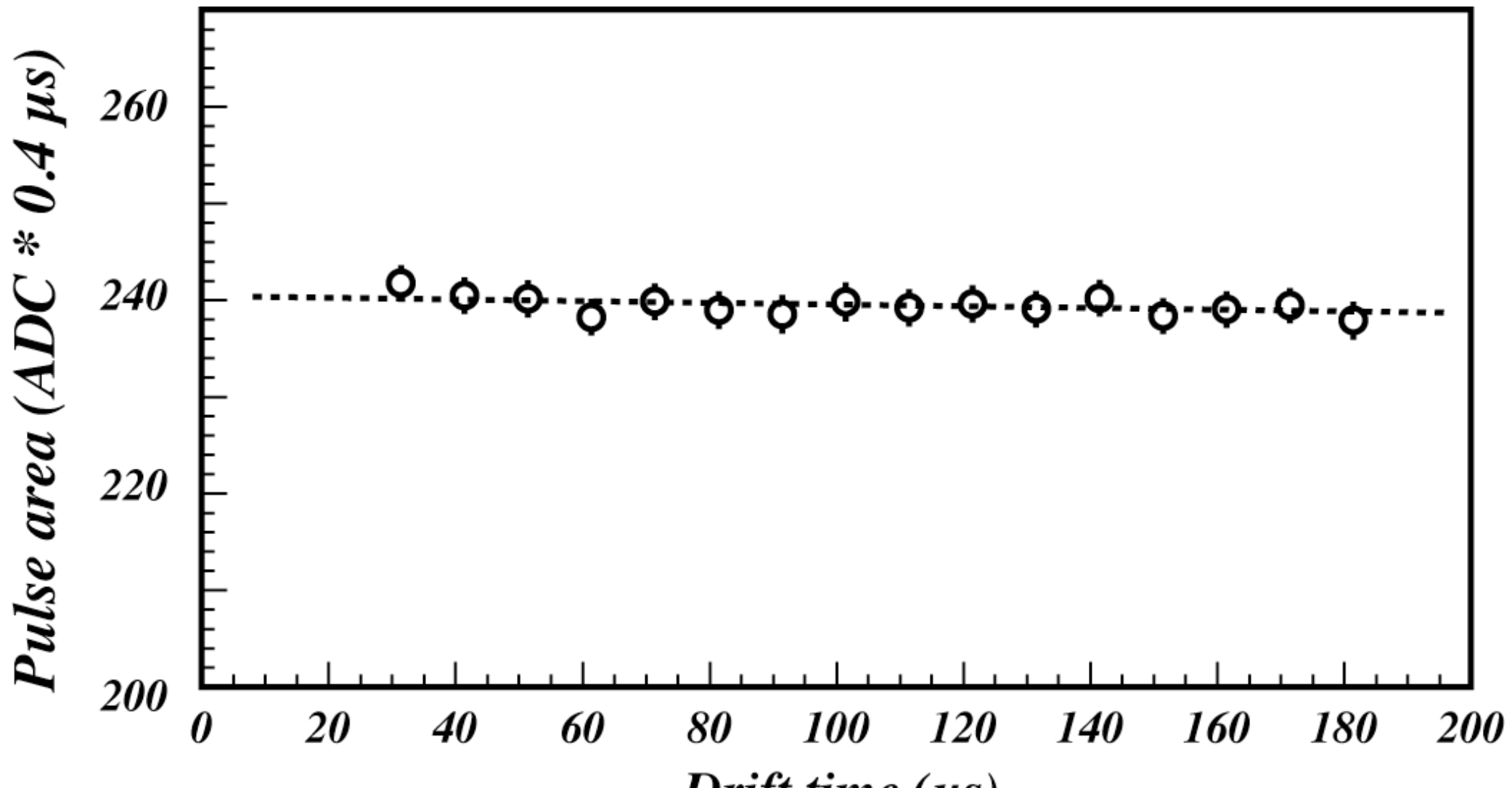

**Figure 9**. Pulse height distribution a function of the hit time for the global event sample. The fitted line of the sample is $1/\tau = (0.047 \pm 0.012)\ \text{ms}^{-1}$ ($\chi^2$ / d.o.f = 11 / 14).

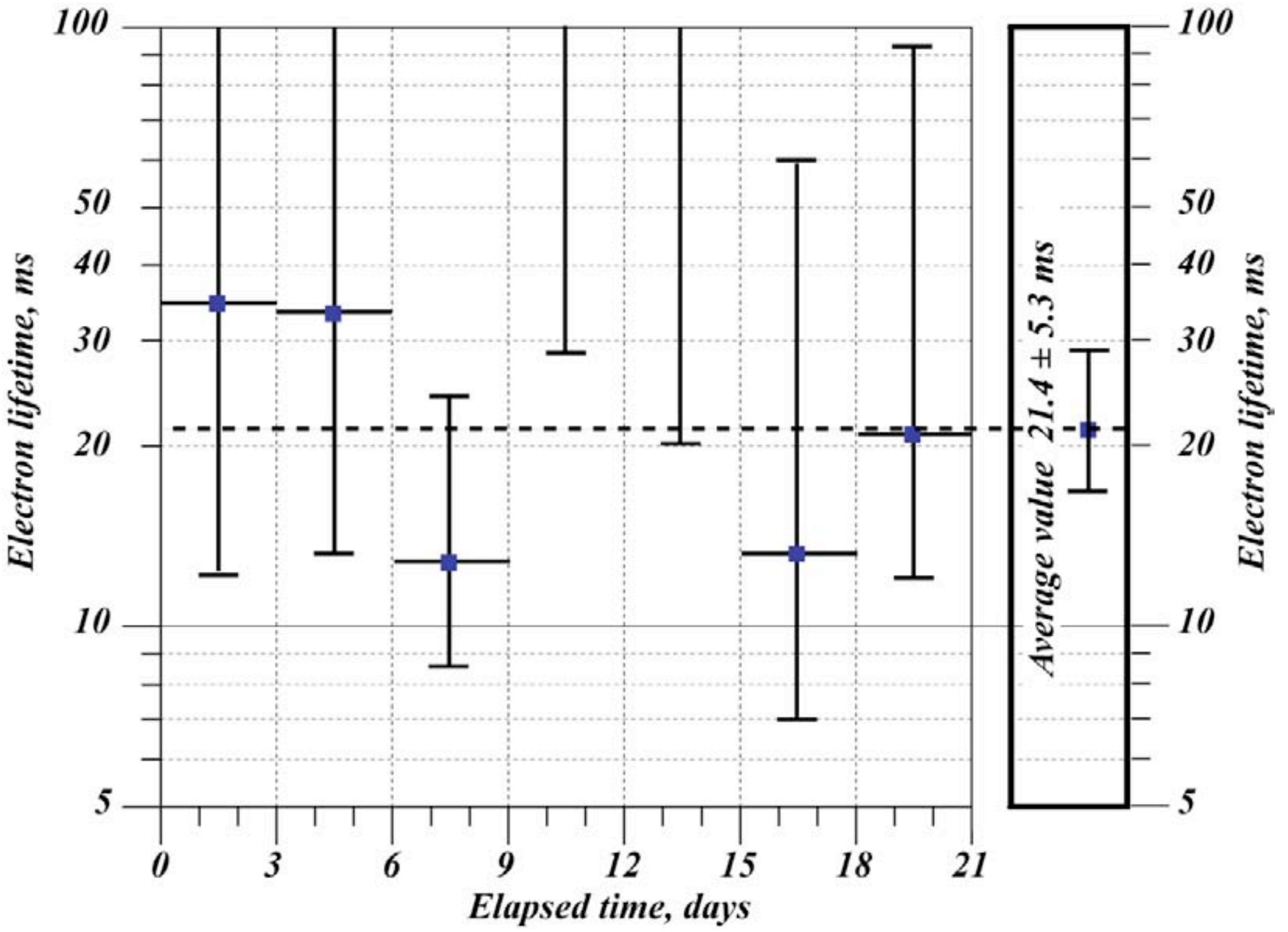

**Figure 10**. The free electron lifetime calculated as a function of the time of the measurement, in bins of 3 days. The over-all result is (21.4 ± 5.3) ms, as shown in Figure 9.